\documentclass[aps,prc]{revtex}
\usepackage[T1]{fontenc}
\usepackage{graphics}

\begin{document}
\draft

\title{Properties of the predicted super-deformed band in $ ^{32}S $.}

\author{R. R. Rodríguez-Guzmán, J.L. Egido and L.M. Robledo}

\address{Departamento de Física Teórica C-XI, Universidad Autónoma de Madrid, 28049
Madrid, Spain.}

\date{\today{}}

\maketitle
\begin{abstract}
Properties like the excitation energy with respect to the ground state, moments
of inertia, $ B(E2) $ transition probabilities and stability against quadrupole
fluctuations at low spin of the predicted superdeformed band of $ ^{32}S $
are studied with the Gogny force D1S using the angular momentum projected generator
coordinate method for the axially symmetric quadrupole moment. The Self Consistent
Cranking method is also used to describe the superdeformed rotational band.
In addition, properties of some collective normal deformed states are discussed.
\end{abstract}

\section{Introduction.}

Recently, a number of papers have addressed the theoretical study of the predicted
super deformed (SD) configuration in the nucleus $ ^{32}S $ by using the
mean field approximation at high spin with several flavors of the Skyrme \cite{Yamagami.99,Moliqe.99}
and also the Gogny interaction \cite{Tanaka.2000}. The interest to study the
SD configuration in $ ^{32}S $ comes from the fact that this SD configuration
is thought to be an intermediate case between the strongly deformed cluster
structures in very light nuclei and the known SD states in the $ A=60 $ region
\cite{Nazarewicz.92,Freer.97}. In fact, it is interesting to understand the
relationship between the predicted SD band in $ ^{32}S $ and the $ ^{16}O+^{16}O $
quasimolecular rotational states observed in this nucleus \cite{Curtis.96,Cindro.88}.
On the other hand, many states up to an excitation energy of around $ 10MeV $
are known experimentally in this nucleus \cite{Brenneisen.97a}. Those states
can be interpreted in terms of the shell model with active particles in the
$ sd $ shell \cite{Brenneisen.97b} and also in terms of the algebraic cluster
model \cite{Cseh.98}. It turns out that some of these states can be interpreted
in terms of deformed intrinsic configurations and, therefore, they can be used
as a test ground to assure the reliability of any interaction meant to describe
the SD band in this nucleus at the mean field level.

The purpose of this paper is to study, using the Gogny interaction \cite{Decharge.80}
with the D1S parameterization \cite{Berger.84}, the properties of the superdeformed
band of the nucleus $ ^{32}S $ focusing on the stability of the superdeformed
minimum at low spin against quadrupole fluctuations. The reason is that in the
theoretical studies mentioned in the above paragraph and also in previous mean
field calculations with the Skyrme \cite{Flocard.84} and Gogny forces \cite{Girod.83}
the SD minimum observed in the energy landscape was very shallow rising serious
doubts about its ability to hold states at low angular momentum when fluctuations
in the quadrupole degree of freedom are taken into consideration (see \cite{Moliqe.99}
for a discussion of this issue). Obviously, at higher spins the rotational energy
makes the SD minimum deeper and therefore much more stable against quadrupole
fluctuations. As a side product of our calculations and also to check the suitability
of the Gogny force for this nucleus we have studied the properties of low lying
normal deformed states and compared them with the available experimental data.
To perform the theoretical analysis, we have used the Angular Momentum Projected
Generator Coordinate Method (AMP-GCM) with the axial quadrupole moment as generating
coordinate and restricted ourselves to $ K=0 $ configurations. The method
allows to obtain an accurate estimate of the excitation energy of the superdeformed
$ 0^{+} $ state with respect to the ground state. The properties of the superdeformed
band obtained with the AMP-GCM are also compared to those of a Self Consistent
Cranking calculation. The choice of the Gogny interaction for this calculation
is backed up not only by the results we obtain for low lying excited states
but also by previous calculations in the context of the Bohr hamiltonian in
the $ \beta  $ and $ \gamma  $ collective variables at zero spin for normal
deformed states that were used to describe neutron and proton pair transfer
reactions \cite{Mermaz.96} and proton scattering \cite{Marechal.99} data with
great success.

\section{Theoretical framework.}

In the framework of the Angular Momentum Projected Generator Coordinate Method
(AMP-GCM) we have used the following ansatz for the $ K=0 $ wave functions
of the system\begin{equation}
\label{eq1}
\left| \Phi ^{I}_{\sigma }\right\rangle =\int dq_{20}f^{I}_{\sigma }(q_{20})\hat{P}^{I}_{00}\left| \varphi (q_{20})\right\rangle .
\end{equation}
 In this expression $ \left| \varphi (q_{20})\right\rangle  $ is the set
of axially symmetric (i.e. $ K=0 $) Hartree-Fock-Bogoliubov (HFB) wave functions
generated with the constraint $ \left\langle \varphi (q_{20})\right| z^{2}-1/2(x^{2}+y^{2})\left| \varphi (q_{20})\right\rangle =q_{20} $
on the mass quadrupole moment. The HFB wave functions have been expanded in
an axially symmetric Harmonic Oscillator (HO) basis with 10 major shells (220
HO states). The two body kinetic energy correction has been fully taken into
account in the variational process. The Coulomb exchange part of the interaction
has not been included into the variational process but added, in a perturbative
fashion, at the end of the calculation. Reflection symmetry has been used as
a selfconsistent symmetry in our HFB wave functions. This is not a real limitation
as octupole instability is expected \cite{Tanaka.2000} at much higher spins
than the ones considered in this work. 

The operator \begin{equation}
\label{eq3}
\hat{P}^{I}_{00}=\frac{2I+1}{8\pi ^{2}}\int d\Omega d^{I}_{00}(\beta )e^{-i\alpha J_{z}}e^{-i\beta J_{y}}e^{-i\gamma J_{z}}
\end{equation}
 appearing in Eq. (\ref{eq1}) is the usual angular momentum projector with
the $ K=0 $ restriction \cite{Hara.95} and $ f^{I}_{\sigma }(q_{20}) $
are the {}``collective wave functions{}'' solution of the Hill-Wheeler (HW)
equation

\begin{equation}
\label{eq2}
\int dq_{20}^{,}{\mathcal H}^{I}(q_{20},q^{,}_{20})
f^{I}_{\sigma }(q_{20}^{,})=
E^{I}_{\sigma }\int dq_{20}^{,}{\mathcal N}^{I}(q_{20},q^{,}_{20})
f^{I}_{\sigma }(q_{20}^{,}).
\end{equation}
In the HW equation we have introduced the projected norm 
${\mathcal N}^{I}(q_{20},q^{,}_{20})=
\left\langle \varphi (q_{20})\right| \hat{P}^{I}_{00}\left| 
\varphi (q_{20}^{,})\right\rangle  $,
and the projected hamiltonian kernel ${\mathcal H}^{I}(q_{20},q^{,}_{20})=
\left\langle \varphi (q_{20})\right| \hat{H}\hat{P}^{I}_{00}\left| 
\varphi (q_{20}^{,})\right\rangle  $. 

The solution of the HW equation for each value of the angular momentum $ I $
determines not only the ground state ($ \sigma =1) $, which is a member of
the Yrast band, but also excited states ($ \sigma =2,3,\ldots  $) that, in
the present context, may correspond to states with different deformation than
the ground state and/or quadrupole vibrational excitations. In order to solve
the HW equation it is usually convenient to work in an orthogonal basis given
by the states $ \left| k^{I}\right\rangle =(n^{I}_{k})^{-1/2}\int dq_{20}u^{I}_{k}(q_{20})\hat{P}^{I}_{00}\left| \varphi (q_{20})\right\rangle  $
defined in terms of the quantities $ u_{k}^{I}(q_{20}) $ and $ n_{k}^{I} $
which are eigenvectors and eigenvalues, respectively, of the projected norm,
i.e. $ \int dq^{,}_{20}{\mathcal N}^{I}(q_{20},q^{,}_{20})u^{I}_{k}(q^{,}_{20})=n^{I}_{k}u^{I}_{k}(q_{20}) $.
The correlated wave functions are written in terms of the new basis as $ \left| \Phi ^{I}_{\sigma }\right\rangle =\sum _{k}g^{\sigma ,I}_{k}\left| k^{I}\right\rangle  $
where the amplitudes $ g^{\sigma ,I}_{k} $ have been introduced. In terms
of these amplitudes it is possible to define {}``collective{}'' wave functions
$ g^{\sigma ,I}(q_{20})=\sum _{k}g_{k}^{\sigma ,I}u_{k}^{I}(q_{20}) $ whose
square, contrary to the $ f_{\sigma }^{I}(q_{20}) $ amplitudes, has the meaning
of a probability. In the solution of the HW equation a technical difficulty
is encountered: for $ q_{20} $ values close to sphericity and $ I\neq 0 $,
the projected norms $ {\mathcal N}^{I}(q_{20},q'_{20}) $ get very small and,
as a consequence, the evaluation of the hamiltonian kernels for those values
of $ q_{20} $, $ q'_{20} $ and $ I $ is prone to strong numerical instabilities.
The most notorious consequence is that the angular momentum projected (AMP)
energy $ E^{I}(q_{20})={\mathcal H}^{I}(q_{20},q_{20})/{\mathcal N}^{I}(q_{20},q_{20}) $
can not be accurately computed for $ q_{20} $ close to sphericity and $ I\neq 0. $
For this reason, whenever the AMP energies are plotted in the next section the
values corresponding to $ q_{20} $ near sphericity will be omitted. However,
this difficulty does not pose any problem for the solution of the HW equation
because the configurations with very small projected norms only contribute to
the orthogonal states $ \left| k^{I}\right\rangle  $ with very small values
of $ n_{k}^{I} $ and these states are not taken into account in the solution
of the HW equation. Let us also mention that details pertaining the evaluation
of the hamiltonian kernels for density dependent forces are given in \cite{Rayner.2000a,Rayner.2000b}.

Finally, it has to be said that one of the drawbacks of the method is that the
intrinsic wave functions are determined before the projection, i.e. we are using
a Projection After Variation (PAV) method. A better way to proceed would be
to use Projection Before the Variation (PBV) \cite{Ring.80} but this would
lead us to a triaxial projection which, for the moment, is extremely costly
for the full configuration spaces used with the Gogny force (see \cite{Schmid.87}
for an implementation of PBV in small configuration spaces). According to the
results of refs. \cite{Villars.71,Friedman.70} the PBV energy for strongly
deformed systems (computed after several approximations) can be written as 
$ E_{PBV}(I)=\langle H\rangle-\frac{<\vec{J}^{2}>}{2{\mathcal J}_{Y}}
+\frac{\hbar ^{2}I(I+1)}{2{\mathcal J}_{TV}} $
where $ {\mathcal J}_{Y} $ and $ {\mathcal J}_{TV} $ are the Yoccoz (Y)
and Thouless-Valatin (TV) moments of inertia, respectively. For the PAV one
has to replace the TV moment of inertia by the Y one in the last term of the
previous formula. These results mean that the rotational energy correction at
$ I=0 $ is already well described in the PAV but for $ I\neq 0 $ one has
to use the PBV method. As discussed in \cite{Rayner.2000b} the effect of the
PBV can be estimated by carrying out Self Consistent Cranking (SCC) calculations.
When the results of these calculations are compared to those of the AMP in several
Mg isotopes it is found that the SCC gamma ray energies are quenched by a factor
0.7 with respect to the AMP ones. As we will see in the next section the same
quenching factor appears for the SD rotational band when the SCC and the AMP
results are compared.

\section{Discussion of the results.}

In Figure 1 we present the results of the HFB calculations used to generate
the intrinsic states $ \left| \varphi (q_{20})\right\rangle  $. On the left
hand side of the figure we show the HFB energy (lower panel) along with the
$ \beta _{4} $ deformation parameter (middle panel) and the particle-particle
energy $ E_{pp}=-1/2Tr(\Delta \kappa ^{*}) $ (upper panel) for protons and
neutrons as a function of the quadrupole moment $ q_{20} $. The energy curve
shows a deformed ground state minimum at $ q_{20}=0.4b $ $ (\beta _{2}=0.19) $
which is only $ 130KeV $ deeper than the spherical configuration. A very
shallow super deformed (SD) minimum at $ q_{20}=1.9b $ $ (\beta _{2}=0.73) $
is also observed at an excitation energy $ E^{HFB}_{x}(SD)=9.85MeV $. To
study the effect of the finite size of the basis in the HFB energies and in
the excitation energy of the SD minimum we have carried out calculations including
$ 18 $ major shells for the HO basis\footnote{%
This basis is considered by some authors \cite{Moliqe.99} as almost indistinguisable
from an infinite basis for the nucleus considered here.
} (1140 states) for both the normal deformed (ND) and SD HFB minima and found that the corresponding
energies are shifted downwards by $ 759KeV $ and $ 1071KeV $ respectively.
As a consequence, the excitation energy of the SD minimum gets reduced by $ 312KeV $
(a $ 4\% $ effect) up to the value $ E^{HFB}_{x}(SD)=9.54MeV $. The hexadecupole
deformation parameter $ \beta _{4} $ is seeing to increase with increasing
quadrupole moments and reach at the SD minimum the rather high value $ \beta _{4}=0.33 $.
Concerning the particle-particle correlation energies $ E_{pp} $ we observe
that their values for protons and neutrons are nearly identical and they go
to zero in both the normal deformed and superdeformed minima. This implies that
dynamical pairing effects could be relevant for the description of both the
ND and SD bands. On the right hand side of the figure we have plotted the matter
density contour plots (at a density $ \rho _{0}=0.08fm^{-3} $) for several
values of $ q_{20} $. Only for $ q_{20} $ values greater or equal 3.2
b the matter density distribution resembles the one corresponding to two touching
$ ^{16}O $ spherical nuclei. On the other hand, the matter distribution corresponding
to the SD minimum ($ q_{20}=2b $) resembles closely the density obtained
in the two-center harmonic oscillator model by coalescing (i.e. taking the distance
between the centers of the two harmonic oscillator potentials equal to zero)
the configuration corresponding to two separate $ ^{16}O $ nuclei in their
ground states (see \cite{Freer.97} and references therein). In this case and
according the Harvey prescription the resulting $ ^{32}S $ nucleus is not
formed in its ground state but rather in the excited configuration $ (0)^{4}(1)^{12}(2)^{12}(3)^{4} $
where four particles have been promoted from the $ N=2 $ major shell to the
$ N=3 $ one. According to the ideas developed by Rae \cite{Rae.88} relating
clustering to the appearance of shell gaps in the single particle spectrum the
above configuration correspond to a deformed nucleus with an axis ratio of $ 2:1 $.
In fact, the matter distribution of the SD minimum has an axis ratio $ z/x=1.92 $,
a proton mean square radius of $ 3.66fm $ and deformation parameters $ \beta _{2}=0.73 $
and $ \beta _{4}=0.33 $. To study further the connection between these ideas
and our HFB results for the SD minimum we have computed the spherical shell
occupancies $ \nu (nlj)=\sum _{m}\left\langle \varphi \right| c^{+}_{nljm}c_{nljm}\left| \varphi \right\rangle  $
for the intrinsic SD wave function. The quantities $ \nu (nlj) $ give the
occupancy (or contents) of the HO orbital $ nlj $ in the intrinsic wave function
$ \left| \varphi \right\rangle  $. The positive parity level occupancies
are 3.168, 1.688 and 0.706 for the $ d_{5/2} $, $ d_{3/2} $ and $ 2s_{1/2} $
orbitals respectively whereas for the negative parity levels the occupancies
are 0.992, 0.282, 0.864 and 0.174 for the $ 1f_{7/2} $, $ 1f_{5/2} $,
$ 2p_{3/2} $ and $ 2p_{1/2} $ orbitals respectively (the quantities for
proton and neutrons are very similar and therefore only the proton values are
given). Therefore we have for the SD intrinsic state 11.124 particles in the
$ N=2 $ major shell and 4.624 in the $ N=3 $ one in good agreement with
the Harvey prescription. As a consequence of these occupancies, we get contributions
to the quadrupole moment both from the positive parity ($ sd) $ orbitals
and the negative parity ($ pf) $ ones. The two contributions turn out to
be nearly the same for the SD intrinsic wave function. These values of the occupancies
also imply that for a proper description of the SD configuration in terms of
the shell model one needs to consider not only the $ sd $ shell but also
the complete $ fp $ shell. Finally, let us mention that the occupancies of
the negative parity orbitals just mentioned are very small for the ND minimum
as expected.

On the left hand side of Figure 2 we have plotted the AMP energy curves for
$ I=0,\ldots ,12\hbar  $ (full lines) along with the HFB energy curve (dashed
line) as a function of the quadrupole moment. The AMP $ I=0\hbar  $ energy
curve shows more pronounced ND and SD minima than the HFB one and they are located
at quadrupole moments $ q_{20}=0.55b $ and $ q_{20}=2.02b $ for the ND
and SD configurations. The excitation energy of the SD minimum with respect
to the ground state for $ I=0\hbar  $ is $ E^{AMP}_{x}(SD)=8.22MeV $ to
be compared with the $ 9.85MeV $ obtained in the HFB calculation. Let us
mention that performing the AMP calculations with $ 18 $ shells is extremely
time consuming and therefore we will just use in this case the $ 312KeV $
shift obtained in the HFB to account for the effect of the finite size of the
basis in the excitation energy of the SD band head. We will also use the $ 759KeV $
shift in the energy of the ground state to estimate the binding energy in the
AMP-GCM calculation. If we take into consideration the $ 312KeV $ shift the
excitation energy of the SD minimum in the AMP case becomes $ E^{AMP}_{x}(SD)=7.91MeV $
to be compared with the $ 9.54MeV $ obtained in the HFB case with 18 shells.
We notice that for increasing spins, the superdeformed minimum gets more and
more pronounced and becomes the ground state at spin $ I=8\hbar  $. This
value is four units lower than the HFB predictions of \cite{Moliqe.99,Tanaka.2000}.
Finally, let us mention that the main effect of considering the PBV moments
of inertia (see section 2) in the AMP energy curves would be the lowering of
the $ I\neq 0 $ curves but the $ I=0 $ reference curve should remain unchanged.
Therefore, we do not expect changes in our prediction of the excitation energy
of the SD band head and the angular momentum for which the SD band becomes the
Yrast band coming from the effects of PBV.

\begin{figure}
{\par\centering \resizebox*{0.9\textwidth}{!}
{\rotatebox{-90}{\includegraphics{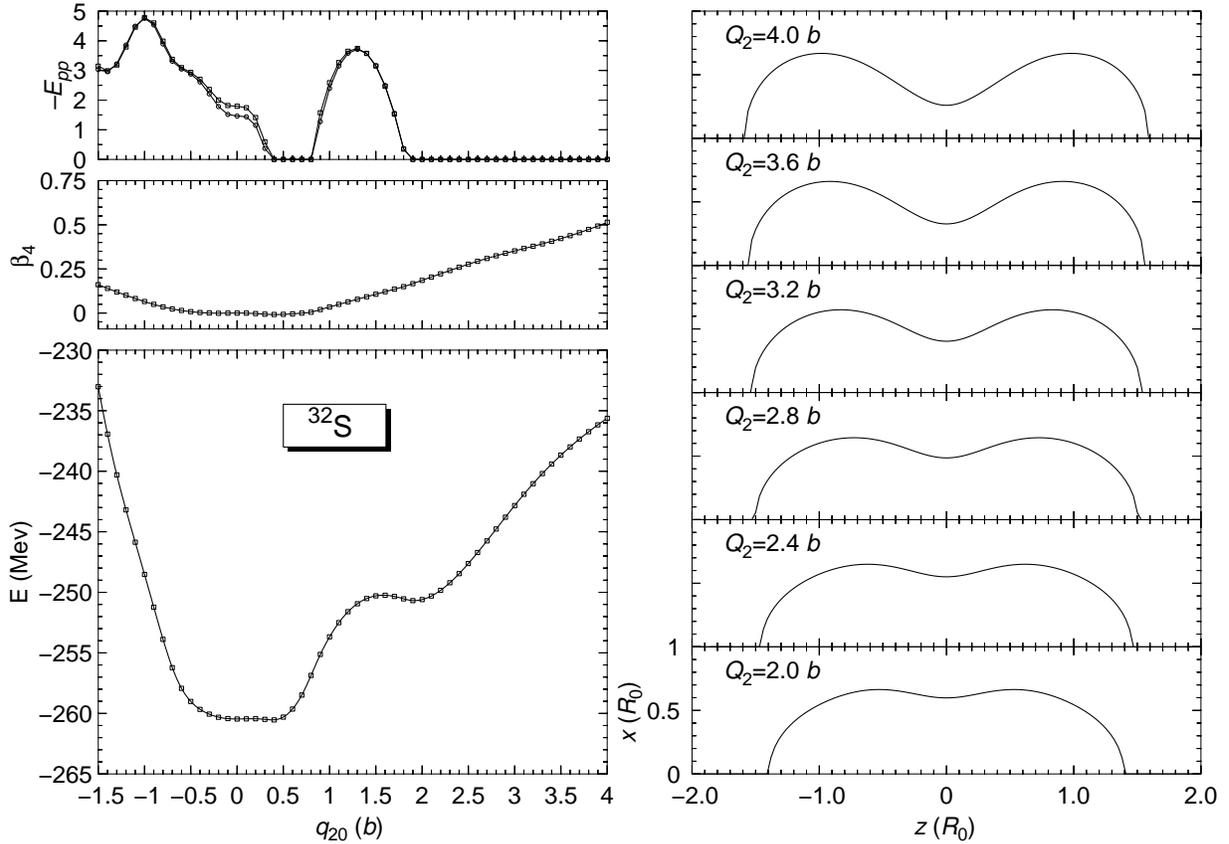}}} \par}
\caption{On the left hand side the HFB energy (lower panel), the \protect$ \beta _{4}\protect $
deformation parameter (middle panel) and the particle-particle correlation energies
\protect$ E_{pp}\protect $ for protons are neutrons (upper panel) are depicted
as a function of the quadrupole moment \protect$ q_{20}\protect $ given in
barns. On the right hand side, contour plots of the matter distribution corresponding
to a density of \protect$ 0.08fm^{-3}\protect $ and different quadrupole
moments are depicted.}
\end{figure}

On the right hand side of Figure 2 we show the energies obtained in the AMP-GCM
calculations for the four lowest-lying solutions of the HW equation (labeled
with the subindex $ \sigma =1,\ldots ,4 $) and spins from zero up to $ 12\hbar  $.
Each level has been placed at a $ q_{20} $ value corresponding to its average
deformation $ \left( \overline{q}_{20}\right) ^{\sigma ,I}=\int dq_{20}\left| g^{\sigma ,I}(q_{20})\right| ^{2}q_{20} $.
The $ I=0\hbar  $ projected energy curve has also been plotted to guide the
eye.
\begin{figure}
{\par\centering \resizebox*{0.7\textwidth}{!}
{\rotatebox{270}{\includegraphics{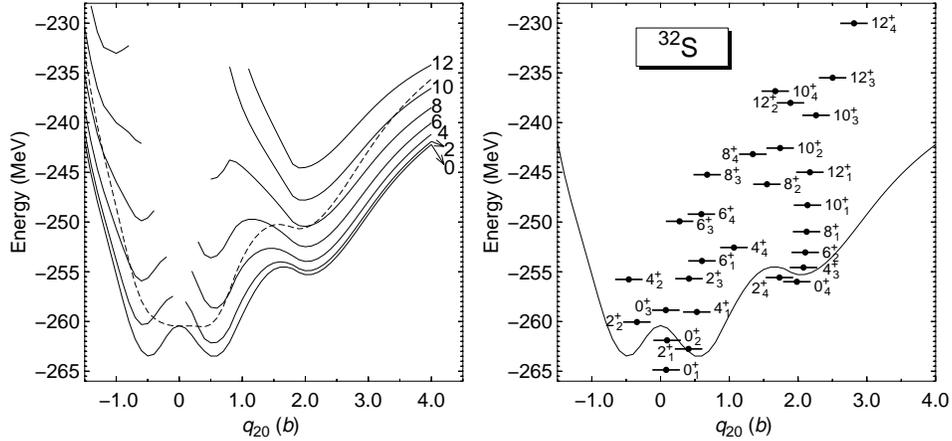}}}\par}
\caption{In the left hand side panel the HFB energy (dashed line) and the angular momentum
projected energies up to \protect$ I=12\hbar \protect $ are plotted as a
function of the quadrupole deformation \protect$ q_{20}\protect $ measured
in barns. See section 2 for an explanation of the missing point around \protect$ q_{20}=0b.\protect $
In the right hand side, the four lowest-lying solutions of the AMP-GCM equation
are plotted for each spin. In both plots the Coulomb exchange energy has not
been added. See text for further details.}
\end{figure}
We first observe that the three lowest $ 0^{+} $ states are spherical whereas
the fourth one is located inside the SD minimum and therefore is the band head
of the SD rotational band. As a consequence of the quadrupole mixing the excitation
energy of the SD $ 0^{+} $ with respect to the ground state increases up
to $ E^{AMP-GCM}_{x}(SD)=8.87MeV $ ($ 8.56MeV $ for the $ 18 $ shells
basis) and we also confirm that the $ 0^{+} $ SD state is stable against
quadrupole fluctuations. We also clearly observe the SD rotational band ($ 0^{+}_{4} $,
$ 2^{+}_{4} $, $ 4^{+}_{3} $, $ 6^{+}_{2} $, $ 8^{+}_{1} $, $ 10^{+}_{1} $and
$ 12^{+}_{1} $) which becomes Yrast at $ I=8\hbar  $ and shows an irregularity
at $ I=2\hbar  $ that is related to the near degeneracy of this SD state
with the $ 2^{+}_{3} $ ND state. We also notice that a bunch of other SD
states with angular momentum $ 10\hbar  $ and $ 12\hbar  $ appear. Let
us also note that, when the Coulomb exchange energy is taken into account the
$ 0^{+}_{1} $ ground state energy becomes $ -270.87MeV $ ($ -271.63MeV $
when the effect of the finite size of the basis is taken into account) in good
agreement with the experimental binding energy of $ 271.780MeV $ \cite{Audi.95}. 

Concerning the normal deformed states, the $ 2_{1}^{+} $, $ 4_{1}^{+} $,
$ 6_{1}^{+} $ and $ 8_{3}^{+} $ states have nearly the same value of $ \overline{q}^{\sigma ,I}_{20} $
and they could be the members of a rotational band with moderate deformation.
Moreover, the $ 2_{2}^{+} $ and $ 4_{2}^{+} $ states are oblate and together
with the $ 0_{2}^{+} $ state could be the member of another moderate deformation
band. Experimentally, there are many known levels at low excitation energy \cite{Brenneisen.97a}
either of positive and negative parity. They can be interpreted in terms of
the Shell Model within the $ sd $ shell \cite{Brenneisen.97b} or the algebraic
cluster model of \cite{Cseh.98}. Some of the levels can be bunched together
as members of $ K=0 $ bands and could be associated to the two bands obtained
in our calculations. In table 1 we compare the available experimental data on
excitation energies and $ B(E2) $ transition probabilities for both $ K=0 $
bands with our predictions. Taking into account that in our calculations we
only take into account the quadrupole degree of freedom and that the Gogny force
has not been fitted for this region of the periodic table we conclude from table
1 that our results are in reasonable agreement with experiment specially for
the lowest lying states. This fact gives us some confidence on the predictions
we make for the SD band. In this respect, we can also mention that we obtain
a spectroscopic quadrupole moment for the $ 2^{+}_{1} $ of $ -13.29fm^{2} $
that compares well with the experimental value \cite{Endt.90} of $ -14.9fm^{2} $.
Our results are also in good agreement with previous calculations with the Gogny
force in the context of the Bohr hamiltonian for the $ \beta  $ and $ \gamma  $
collective variables \cite{Mermaz.96}.
\begin{table}
\vspace{0.3cm}
{\centering \begin{tabular}{|c|c|c|c|c|c||c|c|c|c|c|c|} 
\multicolumn{3}{|c||}{Band 1 (Exp)}&
\multicolumn{3}{c||}{Band 1 (Th) }&
\multicolumn{3}{c||}{Band 2 (Exp) }&
\multicolumn{3}{c|}{Band 2 (Th) }\\
\hline 
\hline 
$ J^{\pi } $&
$ E $&
$ B(E2) $&
$ J^{\pi } $&
$ E $&
$ B(E2) $&
$ J^{\pi } $&
$ E $&
$ B(E2) $&
$ J^{\pi } $&
$ E $&
$ B(E2) $\\
\hline 
\hline 
$ 0^{+} $&
0&
--&
$ 0_{1}^{+} $&
0&
--&
$ 0^{+} $&
3.778&
--&
$ 0_{2}^{+} $&
2.975&
--\\
\hline 
$ 2^{+} $&
2.230&
60$ \pm  $ 6&
$ 2_{1}^{+} $&
2.107&
72.3&
$ 2^{+} $&
4.282&
48.8({*})&
$ 2_{2}^{+} $&
4.816&
58.0\\
\hline 
$ 4^{+} $&
4.459&
72$ \pm  $ 12&
$ 4_{1}^{+} $&
5.825&
119.8&
$ 4^{+} $&
6.852&
35.4$ ^{+18.6}_{-8.4} $&
$ 4_{2}^{+} $&
9.097&
132.2\\
\hline 
$ 6^{+} $&
8.346&
>22.2&
$ 6_{1}^{+} $&
10.962&
142.8&
$ 6^{+} $&
9.783&
39.6({*})&
--&
--&
--
\end{tabular}}
\caption{Experimental and theoretical values for the excitation energies (in MeV) and
\protect$ B(E2,I\rightarrow I-2)\protect $ transition probabilities (in \protect$ e^{2}fm^{4}\protect $
) of the two \protect$ K=0\protect $ bands of \protect$ ^{32}S\protect $.
Values marked with an star correspond to theoretical predictions found in \protect\cite{Cseh.98}.}
\end{table}

Coming back to the SD band we have also carried out SCC-HFB calculations for
the SD intrinsic state to estimate the effect of PBV in the moment of inertia
of the SD band (see sec. 2). 
\begin{figure}
{\par\centering \resizebox*{0.8\textwidth}{!}
{\rotatebox{270}{\includegraphics{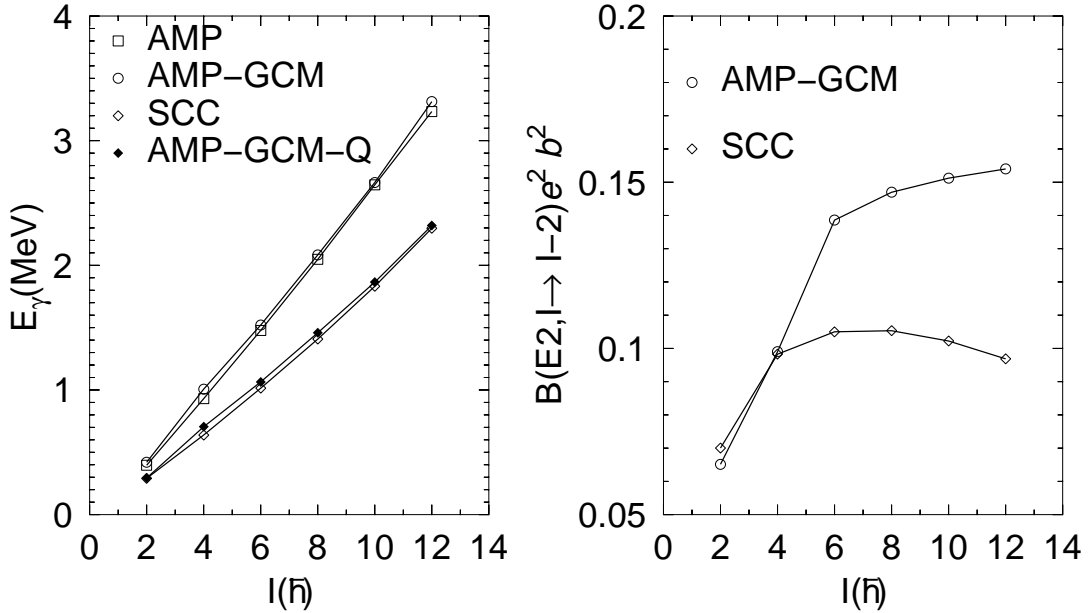}}} \par}

\caption{On the left hand side panel the gamma ray energies \protect$ E_{\gamma }(I)=E(I)-E(I-2)\protect $
in \protect$ MeV\protect $ for the SD configuration are depicted as a function
of spin \protect$ I(\hbar )\protect $ for the AMP, AMP-GCM and SCC calculations.
In addition, the results of the AMP-GCM quenched by a factor 0.7 (AMP-GCM-Q)
are also depicted. In the right hand side panel the theoretical results for
the \protect$ B(E2,I\rightarrow I-2)\protect $ transition probabilities in
units of \protect$ e^{2}b^{2}\protect $ are depicted as a function of spin
for the SCC and AMP-GCM calculations.}
\end{figure}
In the left panel of Figure 3 we have plotted the gamma ray energies $ E_{\gamma }(I)=E(I)-E(I-2) $
as a function of $ I $ for the SD band obtained with different approaches.
Namely, angular momentum projection of the SD intrinsic configuration, the result
of the AMP-GCM and the SCC-HFB results. We observe that the effect of the quadrupole
mixing is very small as the AMP and the AMP-GCM $ E_{\gamma }(I) $ are nearly
identical. However, the SCC-HFB results are clearly different from the AMP ones.
The reason was mentioned before and has to do with the fact that the moments
of inertia computed in the framework of projection before (Thouless-Valatin)
and after (Yoccoz) variation differ considerably. This fact has already been
observed in similar calculations carried out in the same framework for several
$ Mg $ isotopes \cite{Rayner.2000b}. In those calculations we noticed that
the SCC-HFB moment of inertia turned out to be a factor 1.4 bigger than the
AMP one in all the nuclei considered. The same happens here as can be seen by
comparing the SCC-HFB curve with the one labeled AMP-GCM-Q which is obtained
by quenching the AMP-GCM gamma ray energies by a factor 0.7. In the following
we will consider the AMP-GCM-Q results for the SD rotational band as our {}``best{}''
prediction. However, we have to take into account that the pairing correlations
in the SD minimum are negligible and therefore one can expect that dynamic pairing
could play an important role in determining the SCC moment of inertia. The static
moment of inertia obtained in the AMP-GCM-Q (or the SCC-HFB) calculation is
rather constant as a function of spin and has an average value of $ 10.3\hbar ^{2}MeV^{-1} $which
correspond to $ k\equiv \hbar ^{2}/(2{\mathcal J})=48.5KeV $. 

In the right panel of Figure 3 we show the $ B(E2,I\rightarrow I-2) $ transition
probabilities along the SD band for the AMP-GCM calculation and the SCC-HFB
ones computed within the rotational approximation. In both cases, the bare proton
charge has been used. Both transition probabilities agree well up to $ I=6\hbar  $
where the AMP-GCM results become bigger. This is related to the fact that in
the SCC-HFB the Coriolis anti-stretching effect diminishes the intrinsic deformation
from $ \beta _{2}=0.74 $ at $ I=0 $ to $ \beta _{2}=0.68 $ at $ I=12 $
whereas in the AMP-GCM the quadrupole deformation increases with spin (see the
right hand side panel of Fig. 2). The Coriolis anti-stretching effect cannot
be present in a AMP-GCM calculation because the intrinsic states used there
are those computed at zero spin. The computed $ B(E2) $ are highly collective
and correspond to around $ 100W.u. $ for the $ 2^{+}\longrightarrow 0^{+} $
transition.

\section{Concluding remarks.}

As a result of our calculations we can conclude that the predicted superdeformed
band in $ ^{32}S $ is stable at low spin against quadrupole fluctuations
in spite of the shallowness of the HFB SD minimum. The effect of angular momentum
projection and quadrupole mixing reduces the excitation energy of the SD band
head with respect to the ground state to $ 8.87MeV $ ($ 8.56MeV $ if the
effect of the finite size basis is considered), to be compared to the HFB result
of $ 9.85MeV $ ($ 9.54MeV $ for the $ 18 $ shells basis). We have also
found that the SD band becomes Yrast at angular momentum $ I=8\hbar  $ in
our calculations in contrast with the results of several HFB results where it
becomes Yrast at $ I=12\hbar  $. The computed moment of inertia of the SD
band is of the order of $ 10.3\hbar ^{2}MeV^{-1} $ and the $ B(E2) $ transition
probabilities along this band are very collective, exceeding $ 100W.u $.
The SD band head configuration corresponds to the promotion of four particles
from the $ sd $ to the $ fp $ shell. As a consequence, the matter distribution
of the SD configuration corresponds to two coalescent $ ^{16}O $ nuclei in
the sense of the two-center harmonic oscillator shell model in good agreement
with the Harvey prescription.

\section{Acknowledgments.}

One of us (R. R.-G.) kindly acknowledges the financial support received from
the Spanish Instituto de Cooperacion Iberoamericana (ICI). This work has been
supported in part by the DGICyT (Spain) under project PB97/0023.

\end{document}